\documentclass[aps,twocolumn,superscriptaddress,nofootinbib,showpacs]{revtex4-1}

\usepackage[utf8]{inputenc}
\usepackage[T1]{fontenc}
\usepackage{graphicx}
\usepackage{dcolumn}
\usepackage{bm}
\usepackage{amssymb}
\usepackage{amsmath}
\usepackage{multirow}
\usepackage{booktabs}
\usepackage{array}
\usepackage{hyperref}
\hypersetup{hidelinks}
\usepackage{epstopdf}

\newcommand{\Pom}{\mathbb P}
\newcommand{\Odd}{\mathbb O}
\newcommand{\sbar}{\tilde{s}}

\newcommand{\GeV}{\mathrm{GeV}}
\newcommand{\TeV}{\mathrm{TeV}}

\begin{document}

\title{Odderon in the diffractive dip region}

\author{E.~G.~S.~Luna}
\email{luna@if.ufrgs.br}
\affiliation{Instituto de F\'isica, Universidade Federal do Rio Grande do Sul, Caixa Postal 15051, 91501-970, Porto Alegre, Rio Grande do Sul, Brazil}

\author{M.~G.~Ryskin}
\email{ryskin@thd.pnpi.spb.ru}
\affiliation{NRC Kurchatov Institute, Petersburg Nuclear Physics Institute, Gatchina, St.~Petersburg, 188300, Russia}

\author{V.~A.~Khoze}
\email{v.a.khoze@durham.ac.uk}
\affiliation{Institute for Particle Physics Phenomenology, University of Durham, Durham, DH1 3LE, UK}

\begin{abstract}
We examine possible $C$-odd effects in elastic proton-proton and proton-antiproton scattering in the diffractive dip region. Collider data with $\sqrt{s}>500$ GeV are fitted for $0.2<|t|<1~\GeV^2$, using a simple Regge-inspired parametrization of effective $C$-even and $C$-odd amplitudes. The relative phases are not left free, but are fixed by the signature factors of the corresponding even and odd contributions. In this way, the real and imaginary parts of each term are linked by crossing. The $C$-even amplitude is first described by three effective Pomeron-like terms. The $C$-odd sector is then tested with several one-term Odderon-like forms. We find that the fits do not require a stable nonzero $C$-odd term: the odd contribution is either compatible with zero or becomes weakly constrained when we allow extra freedom. We then test whether the conclusion is changed by adding either a fourth Pomeron-like term or a second Odderon-like term. The improvement obtained with the fourth Pomeron-like term is mainly $C$-even, while the second Odderon-like term is not resolved as an independent contribution. These results provide a complementary test of recent determinations of the Odderon amplitude based on scaling and analyticity.
\end{abstract}

\maketitle

\section{Introduction}
\label{sec:intro}

The Odderon is the $C$-odd, or crossing-odd, component of the high-energy elastic scattering amplitude \cite{LN,Ewerz} (see also the recent reviews \cite{PDG,Ryskin}). It is the negative-signature counterpart of the Pomeron, which gives the leading $C$-even contribution to total and elastic cross sections at high energy. There is no general argument which makes this odd component vanish. In perturbative QCD the lowest-order color-singlet exchange with $C=-1$ is made of three gluons in a symmetric color state, with the color indices coupled through the $SU(3)$ tensor $d^{abc}$ \cite{FukugitaKwiecinski,Bartels,KwiecinskiPraszalowicz,JanikWosiek,BLV}.

The odd contribution is expected to be much smaller than the dominant even one. It is therefore best searched for where the leading amplitude is suppressed. The diffractive dip is such a region. There the imaginary part of the dominant even amplitude becomes small, and a relatively small real odd amplitude can produce a visible change in the shape of $d\sigma/dt$.

The first indications of a possible odd component came from the comparison of elastic $pp$ and $\bar p p$ scattering at the CERN ISR. A difference between the two differential cross sections was observed at $\sqrt{s}=53~\GeV$, in the dip region \cite{Breakstone1985}.
The statistics was limited, however, and the energy was not high enough for secondary Reggeon contributions to be safely neglected. The present evidence is mainly based on the comparison of the D0 $\bar p p$ measurement at $\sqrt{s}=1.96~\TeV$ \cite{D02012} with the LHC $pp$ data \cite{TOTEM276}. The $pp$ cross section has a pronounced dip-bump structure, whereas the $\bar p p$ distribution is much flatter. This difference is usually interpreted as evidence for a non-vanishing $C$-odd exchange \cite{D0TOTEM2021}. A recent global analysis of elastic $pp$, $\bar p p$, and $pn$ scattering over a wide energy range was given in Ref.~\cite{Selyugin2024}.

A different analysis was recently made in Ref.~\cite{CorralRoyon}. There, a scaling property of the LHC $pp$ differential cross sections \cite{Scaling,Csorgo2021} was combined with analyticity of the $S$-matrix. For an amplitude of definite signature, the energy dependence then fixes the corresponding phase.
This led to an extraction of a negative-signature amplitude, both in modulus and in phase, and to a prediction for the Tevatron $\bar p p$ cross section.
A related constraint on the Odderon amplitude, obtained in the color-glass-condensate framework, was presented in Ref.~\cite{Roa2026}.

In the present paper we address the same question by a more direct phenomenological procedure. We do not start from a scaling amplitude.\footnote{The phenomenological scaling observed in Ref.~\cite{Scaling} is not known to follow from a general theoretical argument.} Instead, we fit the available collider data in the dip region. The amplitude is written in a Regge-inspired form, with the crossing phases fixed by the signature factors. The question is then whether a stable odd contribution is required once the $C$-even amplitude has sufficient phenomenological freedom. This point is tested in two steps. We first use three effective Pomeron-like terms and enlarge the Odderon-like contribution only within a one-term form. We then add more freedom in two different ways: either by adding a fourth Pomeron-like term, or by adding a second Odderon-like term. 
This shows how well the dip-region structure can be described by an effective $C$-even amplitude, and how much room is left for a separate $C$-odd term.

The elastic amplitude is written as a sum of terms with
power-like energy dependence. Each term is multiplied by the
corresponding signature factor. Thus the energy dependence of
each contribution is specified explicitly, while its phase is
fixed by crossing. The relative phases of the $C$-even and
$C$-odd components are not fitted independently. No additional
phase parameters, such as the phase $\phi^+$ introduced in
Ref.~\cite{CorralRoyon}, are used.\footnote{Another difference
from Ref.~\cite{CorralRoyon} is that we fit not only the $pp$
data, but also the $\bar p p$ data from the Tevatron and from
the $S\bar p pS$ collider.}

The parametrization is used only in the finite kinematic interval
covered by the fits. Its terms are not identified with bare Regge
exchanges. They should be regarded as effective contributions
which describe the elastic amplitude in this restricted range of
energy and momentum transfer.

The paper is organized as follows. In Sec.~\ref{sec:ampl} we
define the amplitude, the signature factors, and the model
variants. In Sec.~\ref{sec:data} we describe the data sets, the
fit procedure, and the results. The conclusions, together with
the comparison with the scaling-and-analyticity analysis and the
comments on possible cut contributions, are given in
Sec.~\ref{sec:conclusion}. The theoretical basis for the fixed
odd trajectory, and the reasons why it is not
used here, is discussed in Appendix~\ref{app:fixedodd}.

\section{Amplitude, signature factors, and model variants}
\label{sec:ampl}

We write the elastic amplitudes as the sum of crossing-even and
crossing-odd components,
\begin{align}
{\cal F}^{pp}(s,t) &=
{\cal F}_{+}(s,t)-{\cal F}_{-}(s,t),
\label{eq:ppamp}
\\
{\cal F}^{\bar p p}(s,t) &=
{\cal F}_{+}(s,t)+{\cal F}_{-}(s,t).
\label{eq:ppbaramp}
\end{align}
In the fits this decomposition is implemented as
\begin{equation}
{\cal F}(s,t)= \sum_{i=1}^{N_{\Pom}}{\cal F}_{\Pom_i}(s,t) +\tau\sum_{j=1}^{N_{\Odd}}{\cal F}_{\Odd_j}(s,t),
\label{eq:fullamp}
\end{equation}
where $\tau=-1$ for $pp$ and $\tau=+1$ for
$\bar p p$.\footnote{This form may be viewed as a
Regge-inspired generalization of the Phillips-Barger
parametrization~\cite{PB},
\[
{\cal F}(s,t)=A(s)\exp(Bt)+\exp(i\phi)\,C(s)\exp(Dt).
\]
Here the energy dependence of the terms is made explicit, and the
relative phase is fixed by the signature factors rather than
introduced as a free parameter.}
We use the normalization
\begin{equation}
\frac{d\sigma}{dt}(s,t)=
\frac{\pi}{s^{2}}\,|{\cal F}(s,t)|^{2}.
\label{eq:norm}
\end{equation}

The even terms are written as
\begin{equation}
{\cal F}_{\Pom_i}(s,t)=
\eta_{\Pom_i}(t)\,
\beta_{\Pom_i}(t)\,
\sbar^{\alpha_{\Pom_i}(t)} ,
\label{eq:pomterm}
\end{equation}
and the odd terms as
\begin{equation}
{\cal F}_{\Odd_j}(s,t)= \eta_{\Odd_j}(t)\, \beta_{\Odd_j}(t)\, \sbar^{\alpha_{\Odd_j}(t)} .
\label{eq:oddterm}
\end{equation}
Here $\sbar=s/s_0$, with $s_0=1~{\rm GeV}^2$.

The signature factors are taken to be
\begin{equation}
\eta_{\Pom}(t)=
-\exp\left[-{i\pi\alpha_{\Pom}(t)\over2}\right],
\qquad C=+1,
\label{eq:sigP}
\end{equation}
and
\begin{equation}
\eta_{\Odd}(t)=
-i\exp\left[-{i\pi\alpha_{\Odd}(t)\over2}\right],
\qquad C=-1.
\label{eq:sigO}
\end{equation}
With this choice the crossing phases of the even and odd
contributions are fixed by their signature. No additional phase
parameters are introduced.

For each Pomeron-like term, we use a linear trajectory,
\begin{equation}
\alpha_{\Pom_i}(t)=
1+\epsilon_{\Pom_i}
+\alpha'_{\Pom_i}t ,
\label{eq:pomtraj}
\end{equation}
and exponential residues,
\begin{equation}
\beta_{\Pom_i}(t)=
\beta_{\Pom_i}(0)\,
\exp(B_{\Pom_i}t).
\label{eq:pomres}
\end{equation}
The parametrization in Eq.~(\ref{eq:fullamp}) is not meant to be
an asymptotic amplitude.\footnote{For example, a Pomeron-like
term with $\alpha_{\Pom_i}(0)=1+\epsilon_{\Pom_i}>1$ would,
if extrapolated to $s\to\infty$, give a power-like growth of
the forward contribution and hence of the corresponding total
cross section, in conflict with the Froissart bound. Here
Eq.~(\ref{eq:fullamp}) is used only as an effective
parametrization in the finite energy range covered by the data.}

The main $C$-even part contains three effective Pomeron-like terms. Fits with only two such terms were also made.
They give poorer descriptions of the data, with larger values of $\chi^2/{\rm DoF}$, and will not be considered further.
Thus the first three models use the same $C$-even amplitude. The odd part is then enlarged step by step.
In model A1 the odd trajectory is fixed,
\begin{equation}
\alpha_{\Odd}(t)=1,
\label{eq:a5traj}
\end{equation}
and the residue is written as
\begin{equation}
\beta_{\Odd}(t)= \beta_{\Odd}(0)\, \exp(B_{\Odd}t), \qquad B_{\Odd}={1\over2}B_{\Pom_1}.
\label{eq:a5res}
\end{equation}
Model A2 differs from A1 only by allowing $B_{\Odd}$ to be fitted independently. In model A3 the odd trajectory is also released,
\begin{equation}
\alpha_{\Odd}(t)= 1+\epsilon_{\Odd} +\alpha'_{\Odd}t ,
\label{eq:a8traj}
\end{equation}
so that $\epsilon_{\Odd}$ controls the intercept $\alpha_{\Odd}(0)$. It remains to check whether this conclusion is changed by adding more freedom to the parametrization. We make this check in two ways. In A4 a fourth effective Pomeron-like term is added to the even amplitude. In A5 the even amplitude is kept as in A3, but a second Odderon-like term is included. The corresponding B1--B5 fits have the same analytic forms as A1--A5. They differ only in the data ensemble used in the fit, as described in Sec.~\ref{sec:data}. Reduced versions, obtained by removing parameters compatible with zero and refitting the remaining ones, will be introduced there as well.

In all fits the slope parameters $B_{\Pom_i}$ and $B_{\Odd_j}$
are constrained to be positive.
The fitted Pomeron intercept parameters are positive in all cases.

The construction should be interpreted phenomenologically. The effective $C$-even terms in Eq.~(\ref{eq:fullamp}) should not be read as separate Regge poles. In the fitted interval, they provide a representation of the even amplitude, and may include pole contributions, cuts and absorptive corrections. The functions $\alpha_{\Pom_i}(t)$ should therefore not be regarded as particle Regge trajectories.

The same qualification is important for the odd part. This is just the parametrization of a $C$-odd amplitude in a limited energy interval. With one or two Odderon-like terms, the fit can at most test an effective $C$-odd contribution. It cannot separate a pole term from a possible Pomeron--Odderon cut. The theoretical basis for the fixed odd trajectory, and the sense in which it is used here, are discussed in Appendix~\ref{app:fixedodd}.

\section{Data sets, fits, and results}
\label{sec:data}

We fit the available collider elastic $pp$ and $\bar p p$
differential cross-section data in the range
\begin{equation}
 \sqrt{s}>500~\GeV,
 \qquad
 0.2<|t|<1~\GeV^2 .
\label{eq:trange}
\end{equation}
The $\bar p p$ data at $\sqrt{s}=546~\GeV$ are taken from
UA4. The low- and intermediate-$|t|$ points are from
Refs.~\cite{UA4lowt,UA4up05}, while the high-$|t|$ part,
which extends through the shoulder region, is from
Ref.~\cite{UA4hight}. The $\bar p p$ data at
$\sqrt{s}=1.8~\TeV$ are taken from CDF and
E710~\cite{CDF_546_1800,E710_1800}, while the
$\sqrt{s}=1.96~\TeV$ sample is taken from the D0
measurement~\cite{D02012}. The $pp$ data at
$\sqrt{s}=2.76,7,8$, and $13~\TeV$ are taken from the
TOTEM measurements~\cite{TOTEM276,TOTEM7a,TOTEM7b,TOTEM8,TOTEM13}.

The interval in Eq.~(\ref{eq:trange}) contains the diffractive
dip and the subsequent bump. The full sample contains 452
points, distributed as shown in Table~\ref{tab:dataset}. The
statistical and systematic errors are added in quadrature.

\begin{table}[t]
\caption{Number of points in the fitted interval.}
\label{tab:dataset}
\begin{ruledtabular}
\begin{tabular}{cc}
$\sqrt{s}$ & Number of points \\
\hline \\ [-0.3cm]
546 GeV & 73 \\
1.8 TeV & 45 \\
1.96 TeV & 17 \\
2.76 TeV & 37 \\
7 TeV & 68 \\
8 TeV & 37 \\
13 TeV & 175 \\
\hline \\ [-0.3cm]
Total & 452
\end{tabular}
\end{ruledtabular}
\end{table}

A practical difficulty is caused by the 13 TeV data
\cite{TOTEM13}. This set contains many more points than the
other high-energy samples and has smaller point-by-point errors.
A standard least-squares fit may therefore be driven mainly by
the 13 TeV measurements. This is not necessarily the most useful
weighting for the present analysis, where the even-odd
separation is tested through the energy dependence of all data
sets.

We therefore use two ensembles. Ensemble A is the
{\it published-error ensemble}. In this ensemble the experimental
errors are used as published. Ensemble B is the 13 TeV
{\it reweighted ensemble}. It contains the same data points, but
the 13 TeV systematic errors are multiplied by a factor $f$
before they are added in quadrature to the statistical errors,
\begin{equation}
 \sigma_{i,\rm eff}^{2}
 =
 \sigma_{i,\rm stat}^{2}
 +(f\sigma_{i,\rm syst})^{2},
 \qquad
 \sqrt{s}=13~\TeV .
\label{eq:berrors}
\end{equation}
This is not meant as a correction of the experimental
uncertainty. It is a diagnostic procedure, used to test whether
the conclusions are controlled by the large statistical weight of
the 13 TeV sample. We have tested several values of $f$, from
$f=3$ to $f=20$. In all cases the conclusions are unchanged.
The results shown below correspond to $f=3$.

With these ensembles defined, we perform global least-squares fits for each model. The minimum is found with MIGRAD.
The quoted errors are local fit errors associated with the behavior of the $\chi^2$ hypersurface around the minimum.
The fit quality is measured by
\[
 \tilde{\chi}^2 =
 {\chi^2_{\min}\over \nu},
 \qquad
 \nu=N_{\rm data}-N_{\rm par},
\]
where $\nu$ is the number of degrees of freedom. For Gaussian
errors, $\chi^2_{\min}$ is to be compared with a $\chi^2$
distribution with $\nu$ degrees of freedom.

When parameter uncertainties are quoted, they are obtained from
the standard $\Delta\chi^2$ criterion. The $90\%$ confidence
region in the space of the fitted parameters is defined by
\[
 \chi^2-\chi^2_{\min}
 \leq
 \Delta\chi^2_{90\%}(N_{\rm par}) .
\]
The number of free parameters, $N_{\rm par}$, depends on the model variant. Since the data sample contains $N_{\rm data}=452$ points, $N_{\rm par}$ is obtained from the degrees of freedom listed in the fit tables through $N_{\rm par}=N_{\rm data}-\nu$.
The corresponding values of $\Delta\chi^2$ used for the $90\%$ confidence regions are given in Table~\ref{tab:deltachi}.

\begin{table}[t]
\caption{Values of $\Delta\chi^2$ used for $90\%$ confidence
regions in the full parameter space.}
\label{tab:deltachi}
\begin{ruledtabular}
\begin{tabular}{ccc}
Model & $N_{\rm par}$ & $\Delta\chi^2_{90\%}$ \\
\hline \\ [-0.3cm]
A1/11, B1/11 & 11 & 17.28 \\
A1, B1, A2/13, B2/13, A3/13, B3/13 & 13 & 19.81 \\
A2, B2 & 14 & 21.06 \\
A3, B3 & 16 & 23.54 \\
A4, B4, A5, B5 & 20 & 28.41
\end{tabular}
\end{ruledtabular}
\end{table}

\begin{table*}[p]
\centering
\caption{Fit parameters and partial fit qualities for the A1/B1 family. In these models the Odderon-like trajectory is fixed to
$\alpha_{\Odd}(t)=1$, and the slope of the odd residue is tied to the first Pomeron-like term. The reduced fits A1/11 and B1/11
are also shown. }
\begin{ruledtabular}
\begin{tabular}{cccccc}
 & {\bf \footnotesize MODEL A1}  & {\bf \footnotesize MODEL B1} & {\bf \footnotesize MODEL A1/11} & {\bf \footnotesize MODEL B1/11}   \\
 \hline \\ [-0.3cm]
$\epsilon_{{\Bbb P}_{1}}$ & 0.18000$\pm$0.00034 & 0.17326$\pm$0.00059 & 0.18306$\pm$0.00030 & 0.17924$\pm$0.00048 \\
$\alpha^{\prime}_{{\Bbb P}_{1}}$ (GeV$^{-2}$) & 0.08489$\pm$0.00065 & 0.0789$\pm$0.0011 & 0.08461$\pm$0.00063 & 0.09498$\pm$0.00098 \\
$\beta_{{\Bbb P}_{1}}(0)$ & -0.05306$\pm$0.00034 & -0.06216$\pm$0.00069 & -0.05045$\pm$0.00031 & -0.06543$\pm$0.00063 \\
$B_{{\Bbb P}_{1}}$ (GeV$^{-2}$) & 1.011$\pm$0.012 & 1.153$\pm$0.020 & 1.021$\pm$0.011 & 0.980$\pm$0.018 \\ 
$\epsilon_{{\Bbb P}_{2}}$ & 0.07663$\pm$0.00020 & 0.07704$\pm$0.00037 & 0.07364$\pm$0.00020 & 0.04712$\pm$0.00039 \\
$\alpha^{\prime}_{{\Bbb P}_{2}}$ (GeV$^{-2}$) & 0.32140$\pm$0.00061 & 0.3159$\pm$0.0010 & 0.31259$\pm$0.00058 & 0.2447$\pm$0.0010 \\
$\beta_{{\Bbb P}_{2}}(0)$ & 2.959$\pm$0.011 & 3.066$\pm$0.021 & 3.074$\pm$0.011 & 3.949$\pm$0.027  \\ 
$B_{{\Bbb P}_{2}}$ (GeV$^{-2}$) & 1.291$\pm$0.011 & 1.394$\pm$0.018 & 1.422$\pm$0.010 & 2.025$\pm$0.018  \\ 
$\epsilon_{{\Bbb P}_{3}}$ & 0.05552$\pm$0.00079 & 0.0584$\pm$0.0020 & 0.06000$\pm$0.00073 & 0.0858$\pm$0.0010 \\
$\alpha^{\prime}_{{\Bbb P}_{3}}$ (GeV$^{-2}$) & {\bf (0.021$\times 10^{-4}$)$\pm$0.078}  & {\bf (0.20$\times 10^{-4}$)$\pm$0.15} & -  & - \\
$\beta_{{\Bbb P}_{3}}(0)$ & 3.696$\pm$0.055 & 3.51$\pm$0.13 & 3.463$\pm$0.048 & 2.754$\pm$0.052  \\ 
$B_{{\Bbb P}_{3}}$ (GeV$^{-2}$) & 11.958$\pm$0.059 & 12.5$\pm$0.14 & 11.866$\pm$0.054 & 11.214$\pm$0.067  \\ 
$\beta_{{\Bbb O}_{1}}(0)$ & {\bf -0.0020$\pm$0.0022} & {\bf -0.0018$\pm$0.0025}  & - & - \\ [0.1cm]
 \hline \\  [-0.3cm]
$\nu$ & 439  & 439 & 441 & 441  \\
$\tilde{\chi}^{2}_{total}$ & 1.88  & 1.54 & 1.88 & 1.53  \\
$\tilde{\chi}^{2}|_{546\, \textnormal{GeV}}$ & 0.55 & 0.55 & 0.56 & 0.59 \\
$\tilde{\chi}^{2}|_{1.8\, \textnormal{TeV}}$ & 1.19 & 1.19 & 1.19 & 1.16 \\
$\tilde{\chi}^{2}|_{1.96\, \textnormal{TeV}}$ & 4.01 & 3.89 & 3.85 & 3.59 \\
$\tilde{\chi}^{2}|_{2.76\, \textnormal{TeV}}$ & 3.37 & 3.40 & 3.41 & 3.50 \\
$\tilde{\chi}^{2}|_{7\, \textnormal{TeV}}$ & 3.84 & 3.80 & 3.87 & 3.71 \\
$\tilde{\chi}^{2}|_{8\, \textnormal{TeV}}$ & 0.82 & 0.81 & 0.80 & 0.81 \\
$\tilde{\chi}^{2}|_{13\, \textnormal{TeV}}$ & 1.42 & 0.58 & 1.42 & 0.61 \\
\end{tabular}
\end{ruledtabular}
\label{tab:a1b1}
\end{table*}

\begin{table*}[p]
\centering
\caption{Fit parameters and partial fit qualities for the A2/B2 family. These fits differ from A1/B1 by allowing the slope $B_{\Odd}$ of the Odderon-like residue to be fitted independently. The reduced fits A2/13 and B2/13 are also shown. }
\begin{ruledtabular}
\begin{tabular}{cccccc}
 & {\bf \footnotesize MODEL A2}  & {\bf \footnotesize MODEL B2} & {\bf \footnotesize MODEL A2/13} & {\bf \footnotesize MODEL B2/13}   \\
 \hline \\ [-0.3cm]
$\epsilon_{{\Bbb P}_{1}}$ & 0.20205$\pm$0.00033 & 0.19830$\pm$0.00043 & 0.19894$\pm$0.00034 & 0.19830$\pm$0.00056 \\
$\alpha^{\prime}_{{\Bbb P}_{1}}$ (GeV$^{-2}$) & 0.10435$\pm$0.00064 & 0.10968$\pm$0.00079 & 0.09987$\pm$0.00065 & 0.1097$\pm$0.0010 \\
$\beta_{{\Bbb P}_{1}}(0)$ & -0.03526$\pm$0.00022 & -0.04279$\pm$0.00034 & -0.03707$\pm$0.00024 & -0.04279$\pm$0.00045 \\
$B_{{\Bbb P}_{1}}$ (GeV$^{-2}$) & 0.648$\pm$0.012 & 0.651$\pm$0.015 & 0.726$\pm$0.012 & 0.651$\pm$0.019 \\ 
$\epsilon_{{\Bbb P}_{2}}$ & 0.07998$\pm$0.00020 & 0.06329$\pm$0.00030 & 0.08100$\pm$0.00020 & 0.06329$\pm$0.00039 \\
$\alpha^{\prime}_{{\Bbb P}_{2}}$ (GeV$^{-2}$) & 0.31719$\pm$0.00060 & 0.27269$\pm$0.00083 & 0.32023$\pm$0.00061 & 0.2727$\pm$0.0011 \\
$\beta_{{\Bbb P}_{2}}(0)$ & 2.724$\pm$0.010 & 3.237$\pm$0.018 & 2.750$\pm$0.010 & 3.237$\pm$0.023  \\ 
$B_{{\Bbb P}_{2}}$ (GeV$^{-2}$) & 1.330$\pm$0.011 & 1.753$\pm$0.015 & 1.328$\pm$0.011 & 1.753$\pm$0.019  \\ 
$\epsilon_{{\Bbb P}_{3}}$ & 0.05055$\pm$0.00075 & 0.07333$\pm$0.00098 & 0.04917$\pm$0.00081 & 0.0733$\pm$0.0012 \\
$\alpha^{\prime}_{{\Bbb P}_{3}}$ (GeV$^{-2}$) & {\bf (0.0064$\times 10^{-1}$)$\pm$0.0036}  & {\bf (0.15$\times 10^{-4}$)$\pm$0.37} & -  & - \\
$\beta_{{\Bbb P}_{3}}(0)$ & 4.129$\pm$0.058 & 3.199$\pm$0.058 & 4.113$\pm$0.062 & 3.200$\pm$0.074  \\ 
$B_{{\Bbb P}_{3}}$ (GeV$^{-2}$) & 11.860$\pm$0.057 & 11.553$\pm$0.068 & 12.024$\pm$0.060 & 11.553$\pm$0.084  \\ 
$\beta_{{\Bbb O}_{1}}(0)$ & 0.34$\pm$0.19 & 0.27$\pm$0.17  & 0.30$\pm$0.19 & 0.27$\pm$0.22 \\
$B_{{\Bbb O}_{1}}$ (GeV$^{-2}$) & 9.0$\pm$1.2 & 8.3$\pm$1.4  & 9.0$\pm$1.4 & 8.3$\pm$1.8 \\ [0.1cm]
\hline \\  [-0.3cm]
$\nu$ & 438  & 438 & 439 & 439  \\
$\tilde{\chi}^{2}_{total}$ & 1.88  & 1.53 & 1.88 & 1.53  \\
$\tilde{\chi}^{2}|_{546\, \textnormal{GeV}}$ & 0.51 & 0.52 & 0.51 & 0.52 \\
$\tilde{\chi}^{2}|_{1.8\, \textnormal{TeV}}$ & 1.20 & 1.19 & 1.21 & 1.19 \\
$\tilde{\chi}^{2}|_{1.96\, \textnormal{TeV}}$ & 3.56 & 3.36 & 3.61 & 3.36 \\
$\tilde{\chi}^{2}|_{2.76\, \textnormal{TeV}}$ & 3.28 & 3.30 & 3.28 & 3.30 \\
$\tilde{\chi}^{2}|_{7\, \textnormal{TeV}}$ & 4.06 & 3.98 & 4.04 & 3.98 \\
$\tilde{\chi}^{2}|_{8\, \textnormal{TeV}}$ & 0.71 & 0.70 & 0.72 & 0.70 \\
$\tilde{\chi}^{2}|_{13\, \textnormal{TeV}}$ & 1.42 & 0.60 & 1.42 & 0.60 \\
\end{tabular}
\end{ruledtabular}
\label{tab:a2b2}
\end{table*}

\begin{figure*}[p]
\centering
\begin{minipage}{0.49\textwidth}
\centering
\includegraphics[width=\textwidth]{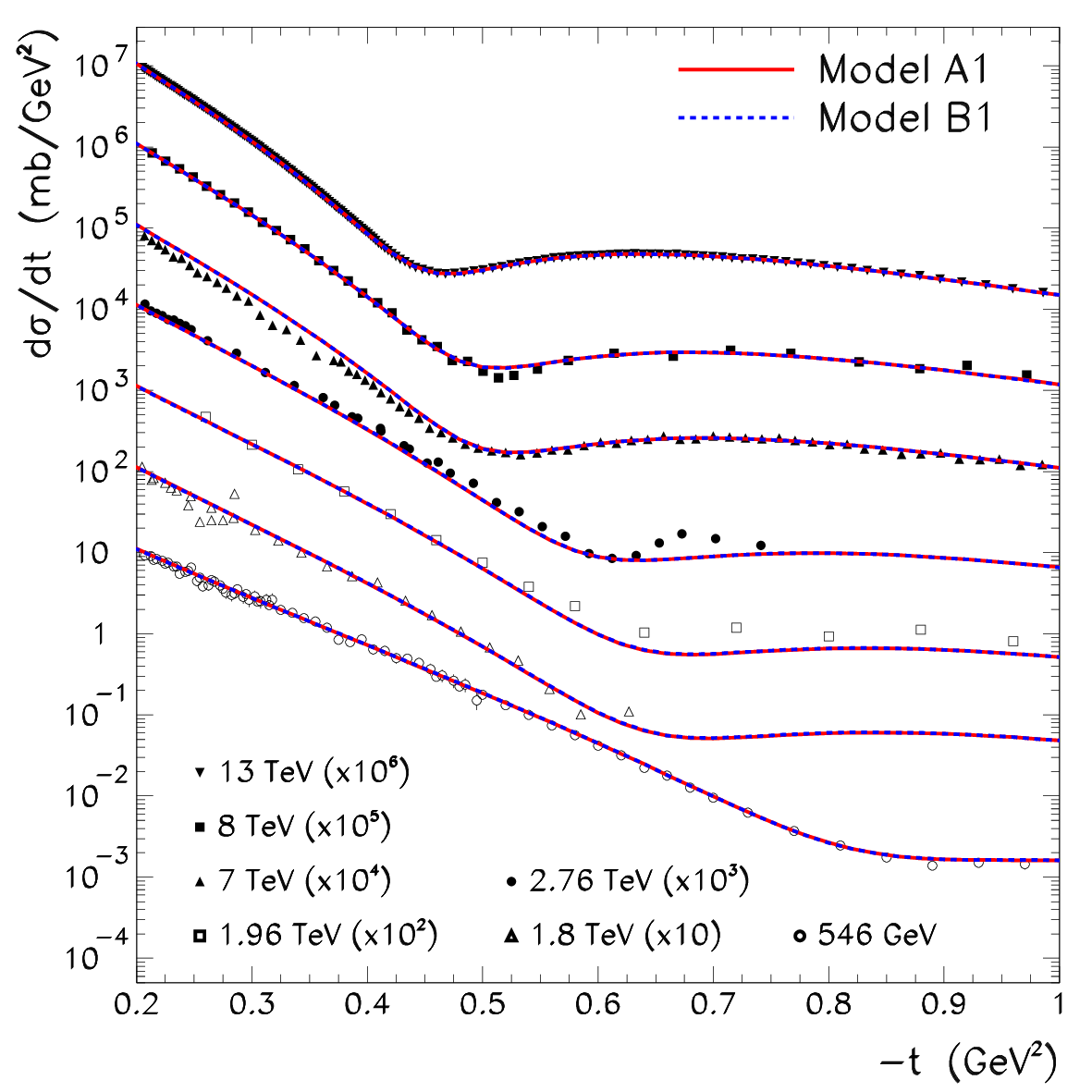}
\end{minipage}
\hfill
\begin{minipage}{0.49\textwidth}
\centering
\includegraphics[width=\textwidth]{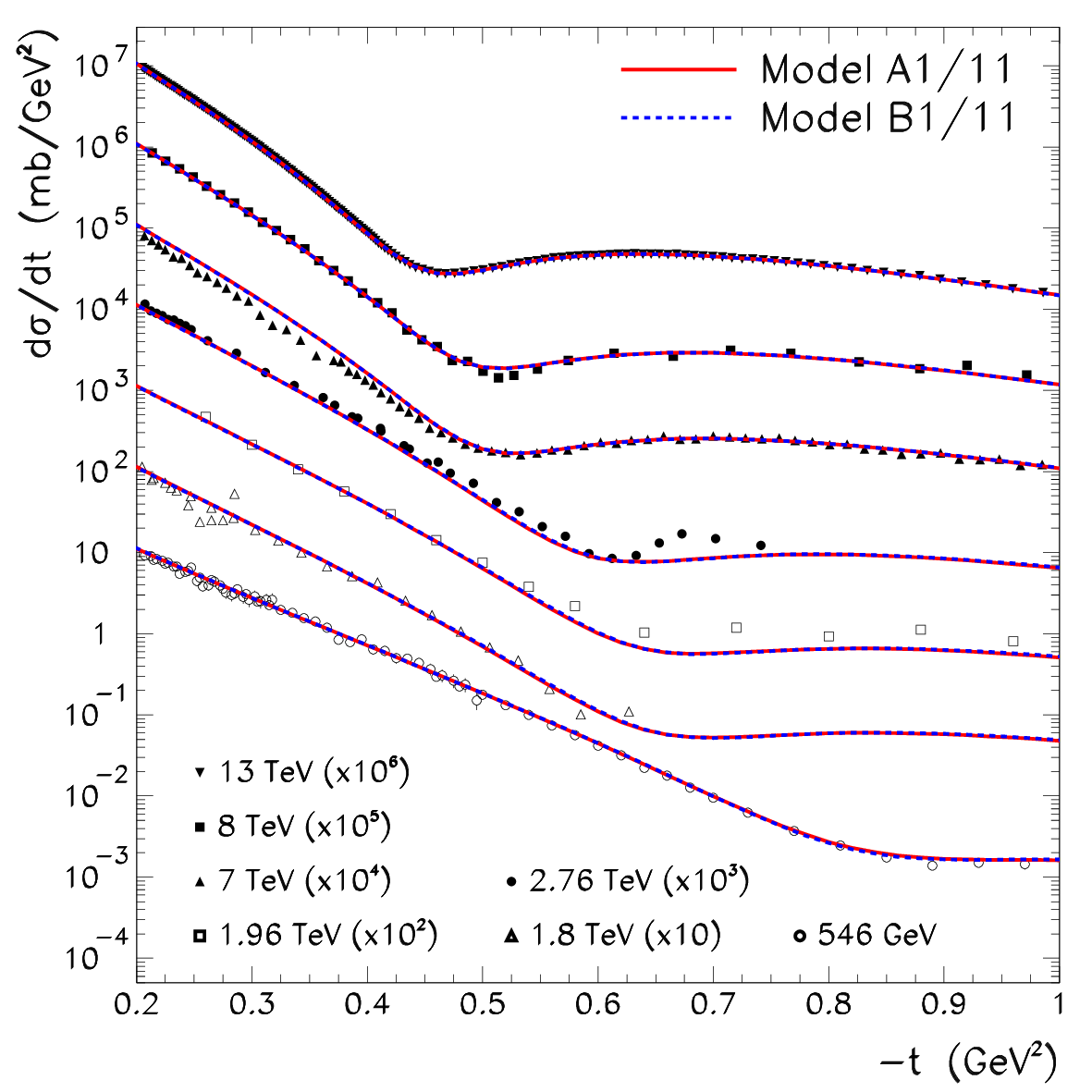}
\end{minipage}
\caption{Differential cross sections for the A1/B1 family.
The left panel shows the full fits A1 and B1. The right panel
shows the reduced fits A1/11 and B1/11.}
\label{fig:lr_a1}
\end{figure*}

\begin{table*}[p]
\centering
\caption{Fit parameters and partial fit qualities for the A3/B3 family. In these fits the Odderon-like intercept and trajectory
slope are also released, with $\alpha_{\Odd}(t)=1+\epsilon_{\Odd} +\alpha'_{\Odd}t$. The reduced fits A3/13 and B3/13 are also shown.}
\begin{ruledtabular}
\begin{tabular}{cccccc}
 & {\bf \footnotesize MODEL A3}  & {\bf \footnotesize MODEL B3} & {\bf \footnotesize MODEL A3/13} & {\bf \footnotesize MODEL B3/13}   \\
 \hline \\ [-0.3cm]
$\epsilon_{{\Bbb P}_{1}}$ & 0.18942$\pm$0.00036 & 0.18559$\pm$0.00048 & 0.18970$\pm$0.00034 & 0.18227$\pm$0.00061 \\
$\alpha^{\prime}_{{\Bbb P}_{1}}$ (GeV$^{-2}$) & 0.086943$\pm$0.00068 & 0.10750$\pm$0.00090 & 0.08823$\pm$0.00066 & 0.0926$\pm$0.0012 \\
$\beta_{{\Bbb P}_{1}}(0)$ & -0.04424$\pm$0.00030 & -0.06719$\pm$0.00060 & -0.04432$\pm$0.00029 & -0.05986$\pm$0.00068 \\
$B_{{\Bbb P}_{1}}$ (GeV$^{-2}$) & 0.968$\pm$0.013 & 0.846$\pm$0.017 & 0.949$\pm$0.012 & 1.003$\pm$0.021 \\ 
$\epsilon_{{\Bbb P}_{2}}$ & 0.07707$\pm$0.00021 & 0.02247$\pm$0.00041 & 0.07609$\pm$0.00020 & 0.05505$\pm$0.00043 \\
$\alpha^{\prime}_{{\Bbb P}_{2}}$ (GeV$^{-2}$) & 0.31557$\pm$0.00063 & 0.1812$\pm$0.0011 & 0.31332$\pm$0.00062 & 0.2579$\pm$0.0012 \\
$\beta_{{\Bbb P}_{2}}(0)$ & 3.018$\pm$0.012 & 4.972$\pm$0.037 & 3.023$\pm$0.011 & 3.722$\pm$0.029  \\ 
$B_{{\Bbb P}_{2}}$ (GeV$^{-2}$) & 1.443$\pm$0.012 & 2.632$\pm$0.020 & 1.450$\pm$0.011 & 1.950$\pm$0.022  \\ 
$\epsilon_{{\Bbb P}_{3}}$ & 0.0429$\pm$0.0010 & 0.09088$\pm$0.00084 & 0.04408$\pm$0.00092 & 0.0716$\pm$0.0015 \\
$\alpha^{\prime}_{{\Bbb P}_{3}}$ (GeV$^{-2}$) & {\bf (0.0051$\times 10^{-1}$)$\pm$0.0065}  & {\bf (0.70$\times 10^{-2}$)$\pm$0.22} & -  & - \\
$\beta_{{\Bbb P}_{3}}(0)$ & 4.429$\pm$0.082 & 2.813$\pm$0.042 & 4.412$\pm$0.075 & 3.300$\pm$0.089  \\ 
$B_{{\Bbb P}_{3}}$ (GeV$^{-2}$) & 12.162$\pm$0.071 & 10.477$\pm$0.052 & 12.080$\pm$0.065 & 11.550$\pm$0.095  \\
$\epsilon_{{\Bbb O}_{1}}$ & {\bf (-0.13$\times 10^{-4}$)$\pm$0.19} & {\bf -0.0084$\pm$0.0091} & - & - \\
$\alpha^{\prime}_{{\Bbb O}_{1}}$ (GeV$^{-2}$) & 0.692$\pm$0.069 & 0.66$\pm$0.12 & 0.695$\pm$0.057 & 0.71$\pm$0.10 \\ 
$\beta_{{\Bbb O}_{1}}(0)$ & 1.95$\pm$0.95 & 2.2$\pm$1.7  & 1.98$\pm$0.91 & 2.2$\pm$1.7 \\
$B_{{\Bbb O}_{1}}$ (GeV$^{-2}$) & {\bf (2.2$\times 10^{-2}$)$\pm$7.4} & {\bf (2.1$\times 10^{-3}$)$\pm$5.0}  & - & - \\ [0.1cm]
\hline \\  [-0.3cm]
$\nu$ & 436  & 436 & 439 & 439  \\
$\tilde{\chi}^{2}_{total}$ & 1.86  & 1.50 & 1.84 & 1.49  \\
$\tilde{\chi}^{2}|_{546\, \textnormal{GeV}}$ & 0.48 & 0.52 & 0.49 & 0.49 \\
$\tilde{\chi}^{2}|_{1.8\, \textnormal{TeV}}$ & 1.19 & 1.24 & 1.19 & 1.18 \\
$\tilde{\chi}^{2}|_{1.96\, \textnormal{TeV}}$ & 3.59 & 3.20 & 3.58 & 3.38 \\
$\tilde{\chi}^{2}|_{2.76\, \textnormal{TeV}}$ & 2.23 & 2.08 & 2.23 & 2.16 \\
$\tilde{\chi}^{2}|_{7\, \textnormal{TeV}}$ & 4.46 & 4.30 & 4.46 & 4.36 \\
$\tilde{\chi}^{2}|_{8\, \textnormal{TeV}}$ & 0.69 & 0.69 & 0.69 & 0.67 \\
$\tilde{\chi}^{2}|_{13\, \textnormal{TeV}}$ & 1.42 & 0.62 & 1.42 & 0.61 \\
\end{tabular}
\end{ruledtabular}
\label{tab:a3b3}
\end{table*}

\begin{figure*}[!t]
\centering
\begin{minipage}{0.49\textwidth}
\centering
\includegraphics[width=\textwidth]{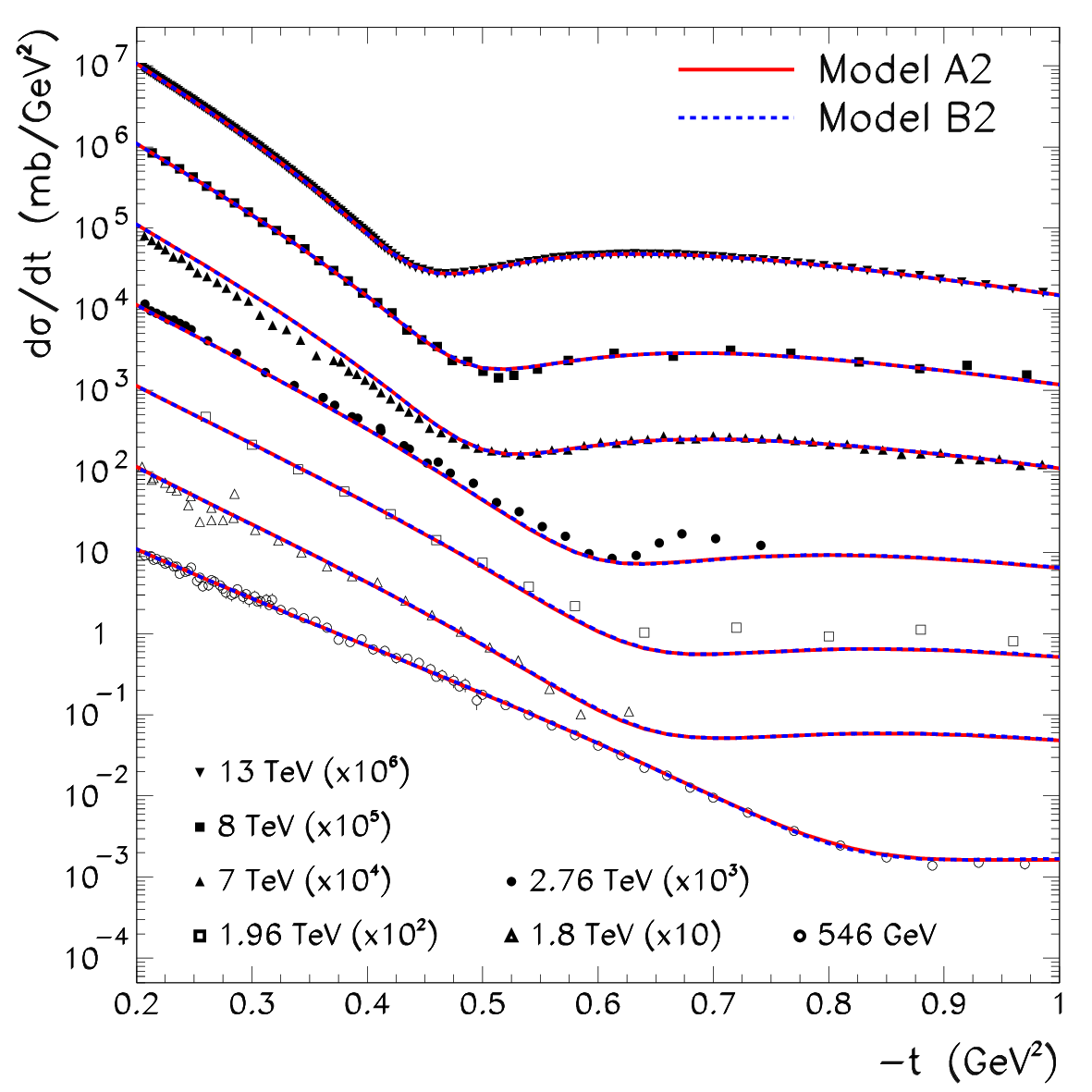}
\end{minipage}
\hfill
\begin{minipage}{0.49\textwidth}
\centering
\includegraphics[width=\textwidth]{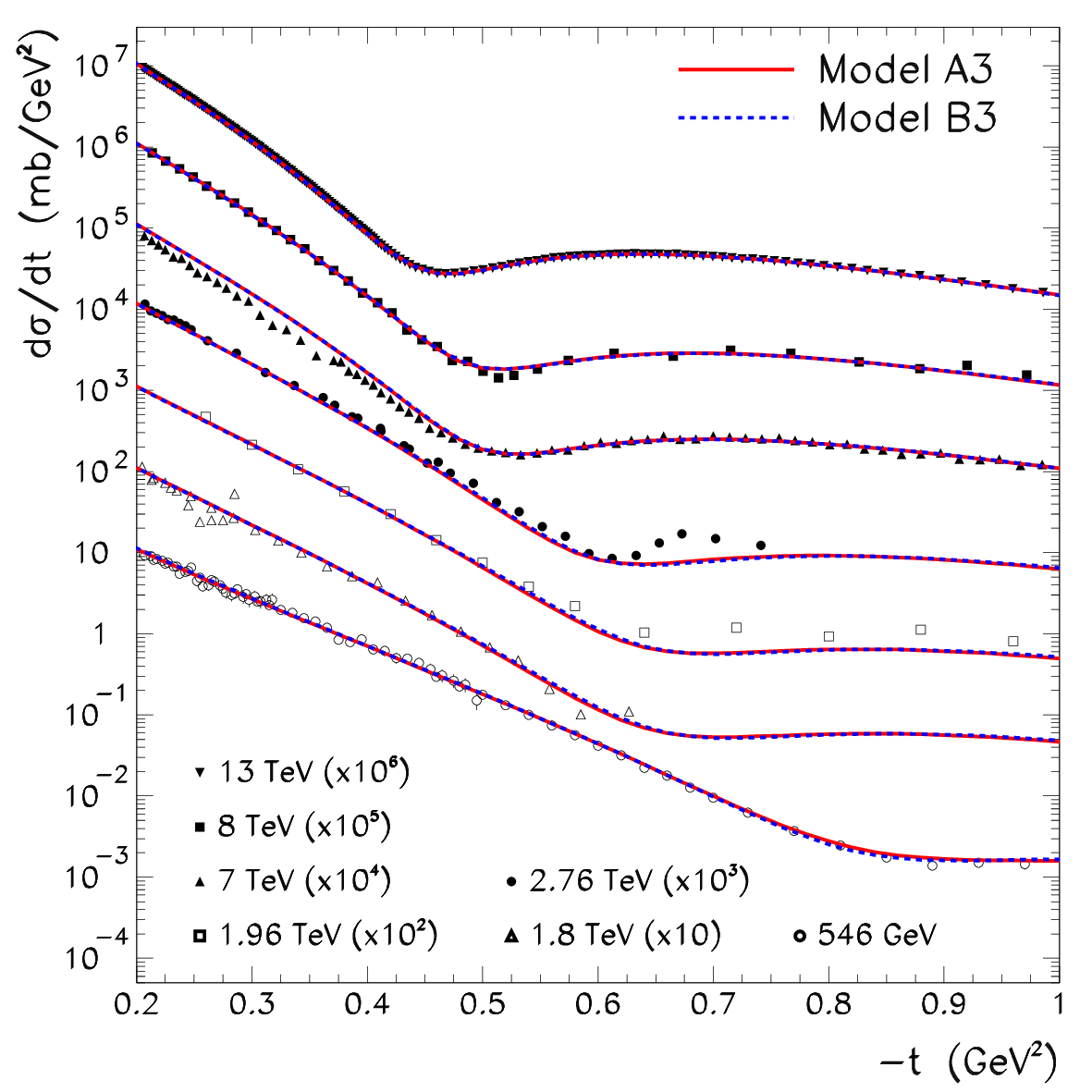}
\end{minipage}
\caption{Differential cross sections for the A2/B2 and A3/B3
families. The left panel shows the A2 and B2 fits, in which the
slope $B_{\Odd}$ of the Odderon-like residue is fitted
independently. The right panel shows the A3 and B3 fits, in which
the Odderon-like intercept and trajectory slope are also fitted.
The corresponding reduced curves are not shown, since they are
indistinguishable from the full-fit curves on this scale.}
\label{fig:lr_a2a3}
\end{figure*}

\begin{table*}[!t]
\centering
\caption{Fit parameters and partial fit qualities for the A4/B4 and A5/B5 families. The A4/B4 fits contain four effective Pomeron-like terms and one Odderon-like term. The A5/B5 fits contain three effective Pomeron-like terms and two Odderon-like terms.}
\begin{ruledtabular}
\begin{tabular}{cccccc}
 & {\bf \footnotesize MODEL A4}  & {\bf \footnotesize MODEL B4} & {\bf \footnotesize MODEL A5} & {\bf \footnotesize MODEL B5}   \\
 \hline \\ [-0.3cm]
$\epsilon_{{\Bbb P}_{1}}$ & 0.10502$\pm$0.00045 & 0.09756$\pm$0.00094 & 0.18890$\pm$0.00023 & 0.18950$\pm$0.00064 \\
$\alpha^{\prime}_{{\Bbb P}_{1}}$ (GeV$^{-2}$) & 0.05196$\pm$0.00080 & 0.0423$\pm$0.0016 & 0.08976$\pm$0.00043 & 0.0940$\pm$0.0012 \\
$\beta_{{\Bbb P}_{1}}(0)$ & -0.1488$\pm$0.0012 & -0.1380$\pm$0.0024 & -0.44870$\pm$0.00019 & -0.04635$\pm$0.00056 \\
$B_{{\Bbb P}_{1}}$ (GeV$^{-2}$) & 1.372$\pm$0.015 & 1.407$\pm$0.030 & 0.9175$\pm$0.0082 & 0.876$\pm$0.022 \\ 
$\epsilon_{{\Bbb P}_{2}}$ & 0.07095$\pm$0.00010 & 0.07094$\pm$0.00019 & 0.07127$\pm$0.00013 & 0.07136$\pm$0.00042 \\
$\alpha^{\prime}_{{\Bbb P}_{2}}$ (GeV$^{-2}$) & 0.18431$\pm$0.00030 & 0.18755$\pm$0.00054 & 0.30816$\pm$0.00040 & 0.3010$\pm$0.0012 \\
$\beta_{{\Bbb P}_{2}}(0)$ & 5.1283$\pm$0.0091 & 5.321$\pm$0.018 & 3.3994$\pm$0.0082 & 3.216$\pm$0.024  \\ 
$B_{{\Bbb P}_{2}}$ (GeV$^{-2}$) & 2.8531$\pm$0.0054 & 2.8963$\pm$0.0097 & 1.5797$\pm$0.0074 & 1.559$\pm$0.021  \\ 
$\epsilon_{{\Bbb P}_{3}}$ & 0.04273$\pm$0.00067 & 0.0460$\pm$0.0020 & 0.07190$\pm$0.00071 & 0.0620$\pm$0.0021 \\
$\alpha^{\prime}_{{\Bbb P}_{3}}$ (GeV$^{-2}$) & 0.0130$\pm$0.0026  & 0.0632$\pm$0.0077 & {\bf (0.088$\times 10^{-3}$)$\pm$0.045}  & {\bf 0.0041$\pm$0.0077} \\
$\beta_{{\Bbb P}_{3}}(0)$ & 5.430$\pm$0.068 & 4.85$\pm$0.18 & 2.689$\pm$0.035 & 3.46$\pm$0.13  \\ 
$B_{{\Bbb P}_{3}}$ (GeV$^{-2}$) & 11.483$\pm$0.048 & 11.57$\pm$0.14 & 12.497$\pm$0.049 & 12.23$\pm$0.14  \\
$\epsilon_{{\Bbb P}_{4}}$ & 0.07834$\pm$0.00016 & 0.07635$\pm$0.00031 & - & - \\
$\alpha^{\prime}_{{\Bbb P}_{4}}$ (GeV$^{-2}$) & 0.01226$\pm$0.00046  & 0.00615$\pm$0.00083 & -  & - \\
$\beta_{{\Bbb P}_{4}}(0)$ & -2.0991$\pm$0.0062 & -2.068$\pm$0.012 & - & -  \\ 
$B_{{\Bbb P}_{4}}$ (GeV$^{-2}$) & 5.1964$\pm$0.0086 & 5.155$\pm$0.015 & - & -  \\
$\epsilon_{{\Bbb O}_{1}}$ & -0.120$\pm$0.044 & -0.129$\pm$0.076 & -0.064$\pm$0.016 & {\bf (-0.43$\times 10^{-2}$)$\pm$0.14} \\
$\alpha^{\prime}_{{\Bbb O}_{1}}$ (GeV$^{-2}$) & 0.59$\pm$0.10 & 0.59$\pm$0.18 & 0.718$\pm$0.047 & 0.783$\pm$0.077 \\ 
$\beta_{{\Bbb O}_{1}}(0)$ & 5.3$\pm$4.5 & 6.3$\pm$3.5  & 10.2$\pm$2.2 & 9.1$\pm$3.9 \\
$B_{{\Bbb O}_{1}}$ (GeV$^{-2}$) & {\bf (84$\times 10^{-7}$)$\pm$27} & {\bf (19$\times 10^{-7}$)$\pm$18}  & {\bf 0.29$\pm$0.79} & {\bf (22$\times 10^{-5}$)$\pm$30} \\
$\epsilon_{{\Bbb O}_{2}}$ & - & - & {\bf (-0.91$\times 10^{-3}$)$\pm$0.19} & {\bf (-0.10$\times 10^{-2}$)$\pm$0.20} \\
$\alpha^{\prime}_{{\Bbb O}_{2}}$ (GeV$^{-2}$) & - & - & 1.00$\pm$0.13 & 0.82$\pm$0.19 \\ 
$\beta_{{\Bbb O}_{2}}(0)$ & - & -  & -9.6$\pm$3.6 & -10.7$\pm$6.0 \\
$B_{{\Bbb O}_{2}}$ (GeV$^{-2}$) & - & -  & 2.2$\pm$1.8 & {\bf 2.0$\pm$2.5} \\ [0.1cm] 
\hline \\  [-0.3cm]
$\nu$ & 432  & 432 & 432 & 432  \\
$\tilde{\chi}^{2}_{total}$ & 1.79  & 1.44 & 1.84 & 1.51  \\
$\tilde{\chi}^{2}|_{546\, \textnormal{GeV}}$ & 0.49 & 0.48 & 0.51 & 0.54 \\
$\tilde{\chi}^{2}|_{1.8\, \textnormal{TeV}}$ & 1.16 & 1.20 & 1.20 & 1.16 \\
$\tilde{\chi}^{2}|_{1.96\, \textnormal{TeV}}$ & 2.91 & 2.93 & 3.37 & 3.29 \\
$\tilde{\chi}^{2}|_{2.76\, \textnormal{TeV}}$ & 1.90 & 1.85 & 2.08 & 2.06 \\
$\tilde{\chi}^{2}|_{7\, \textnormal{TeV}}$ & 4.27 & 4.26 & 4.37 & 4.42 \\
$\tilde{\chi}^{2}|_{8\, \textnormal{TeV}}$ & 0.73 & 0.73 & 0.68 & 0.66 \\
$\tilde{\chi}^{2}|_{13\, \textnormal{TeV}}$ & 1.41 & 0.57 & 1.41 & 0.59 \\
\end{tabular}
\end{ruledtabular}
\label{tab:a4a5}
\end{table*}

\begin{figure*}[!t]
\centering
\begin{minipage}{0.49\textwidth}
\centering
\includegraphics[width=\textwidth]{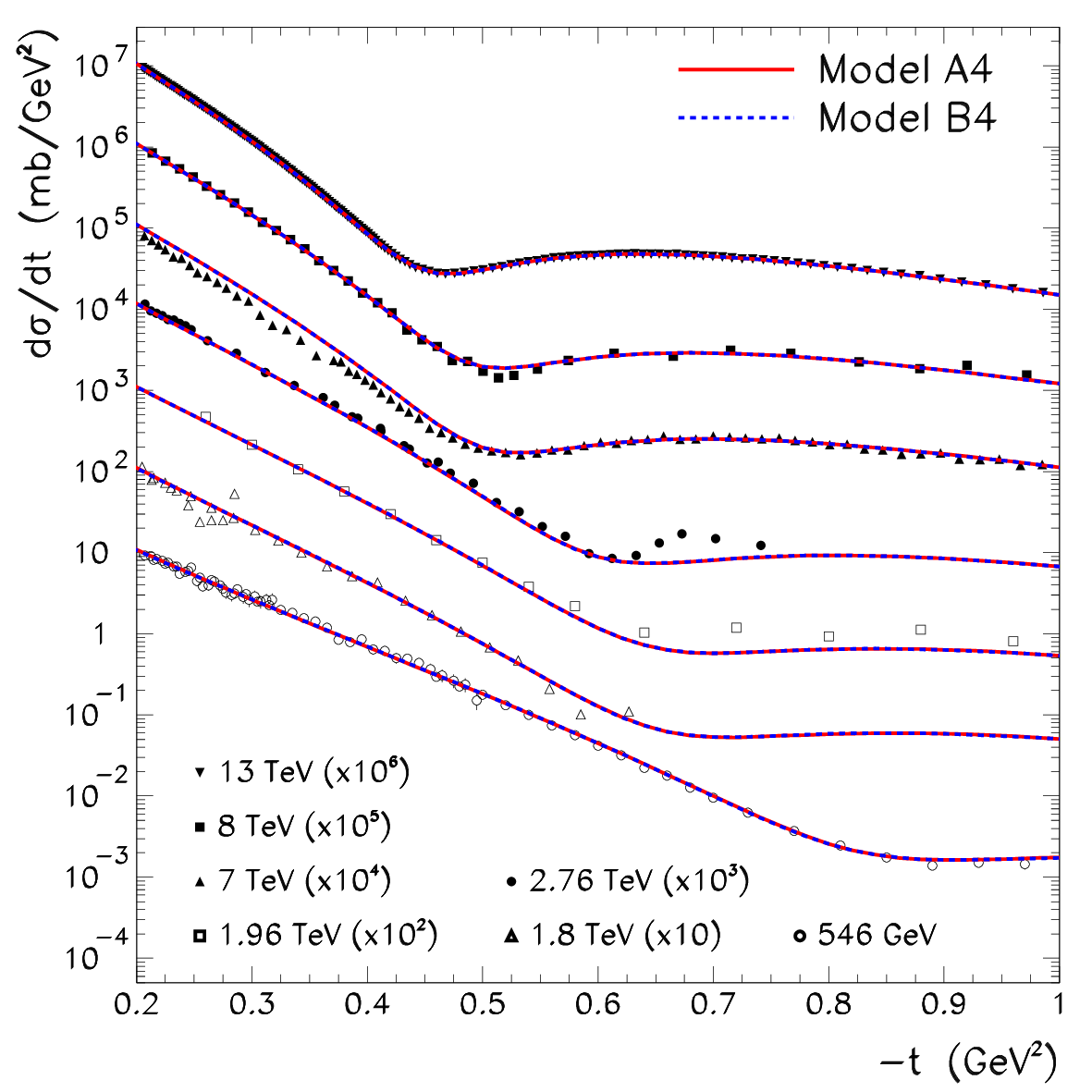}
\end{minipage}
\hfill
\begin{minipage}{0.49\textwidth}
\centering
\includegraphics[width=\textwidth]{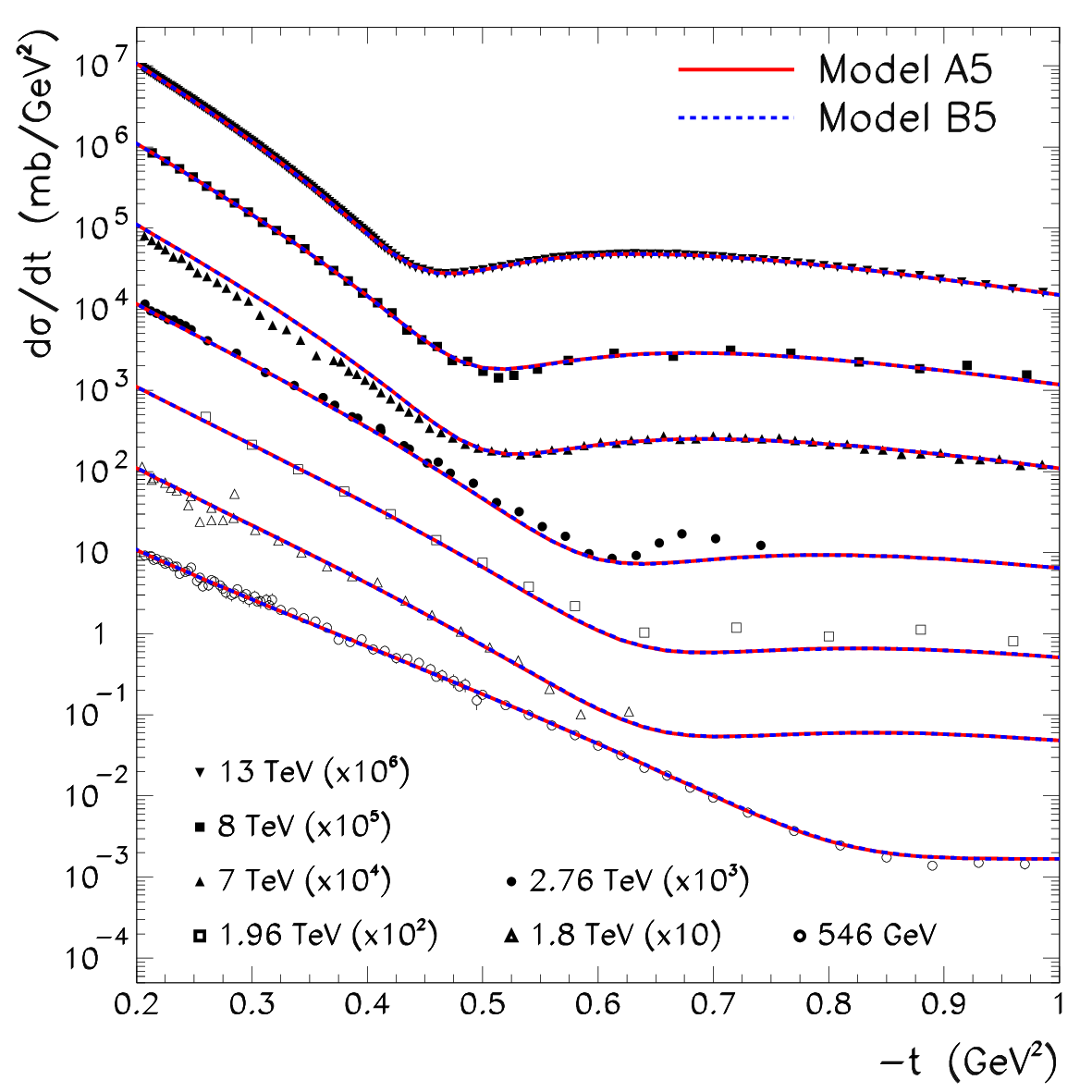}
\end{minipage}
\caption{Differential cross sections for the A4/B4 and A5/B5 families. The left panel shows the A4 and B4 fits, in which a fourth effective Pomeron-like term is added. The right panel shows the A5 and B5 fits, in which a second Odderon-like term is added.}
\label{fig:lr_a4a5}
\end{figure*}

We also perform reduced fits. These are obtained from the
corresponding full fits by removing parameters which are not
effectively constrained by the data. More precisely, we remove
parameters which are compatible with zero, or which are
associated with nearly flat directions of the $\chi^2$
hypersurface. The remaining parameters are then fitted again.
Thus, the reduced fits are not meant to define new physical
models. They are used only as stability tests, to check whether
the quality of the description depends on numerically inactive
parameters.
In the tables, parameters printed in boldface denote quantities that are either compatible with zero within the quoted errors or associated with poorly constrained directions of the fit.

The notation is as follows. A label such as A1 denotes model 1
fitted to the published-error ensemble. A label such as B1 denotes
the same model fitted to the 13 TeV reweighted ensemble. When a
slash is added, the number after the slash gives the number of
free parameters kept in the reduced fit. Thus A1/11 is the
reduced version of A1 with eleven free parameters, while B1/11 is
the corresponding reduced version of B1.

For example, in A1 the slope of the third Pomeron-like trajectory
and the coupling of the Odderon-like term are compatible with
zero. In A1/11 these two parameters are removed, and the
remaining eleven parameters are refitted. The same procedure is
used for the other reduced fits, such as A2/13 and A3/13. This
provides a direct check that the fit quality is not driven by
parameters which are present in the parametrization but not
actually determined by the data.

In the figures the reduced curves are not always shown separately.
Figure~\ref{fig:lr_a1} displays both the full and reduced A1/B1 fits, in order to show explicitly that the removal of the inactive parameters produces no visible change in the curves.
The same happens for the A2/B2 and A3/B3 families. For this reason
Fig.~\ref{fig:lr_a2a3} shows only the corresponding full fits; the
reduced curves would be indistinguishable on the scale of the plot.

The numerical results are collected in Tables~\ref{tab:a1b1}--\ref{tab:a4a5}. Tables~\ref{tab:a1b1}--\ref{tab:a3b3} give the main sequence of fits and the corresponding reduced versions. The first two columns give the fits to the published-error and 13 TeV reweighted ensembles. The last two columns give the corresponding reduced fits. Table~\ref{tab:a4a5} gives the two enlarged variants, A4/B4 and A5/B5. The partial quantities $\chi^2_E/N_E$, quoted for each energy, are evaluated at the global minimum. They are not obtained from independent fits and are used only as diagnostics of the contribution of each data set.

The A1/B1 family is shown in Table~\ref{tab:a1b1} and in
Fig.~\ref{fig:lr_a1}. In these fits the Odderon-like trajectory
is fixed to $\alpha_\Odd(t)=1$, and the slope of the odd
residue is tied to the first Pomeron-like term. The full fits
A1 and B1 already give a reasonable global description of the
data. Their reduced versions, A1/11 and B1/11, are practically
indistinguishable from them. This is shown directly in
Fig.~\ref{fig:lr_a1}.

The A2/B2 family is shown in Table~\ref{tab:a2b2} and in the
left panel of Fig.~\ref{fig:lr_a2a3}. These fits differ from
A1/B1 by allowing the slope $B_\Odd$ of the Odderon-like
residue to be fitted independently. The fitted value of $\beta_\Odd(0)$ is then
nonzero, but the slope is large,
\[
 B_\Odd\simeq 8\hbox{--}9~\GeV^{-2} .
\]
The odd contribution is therefore rapidly suppressed as
$|t|$ increases. At $|t|\simeq 0.5~\GeV^2$, the residue
contains a factor of order
\[
 \exp(-4)\hbox{--}\exp(-4.5),
\]
so that the corresponding term is small in the dip region.

The A3/B3 family is shown in Table~\ref{tab:a3b3} and in the right panel of Fig.~\ref{fig:lr_a2a3}. In these fits the intercept and the trajectory slope of the Odderon-like term are also released.
The intercept parameter remains compatible with zero, so that $\alpha_\Odd(0)$ is compatible with one. The slope of the odd residue is also compatible with zero, but with a large uncertainty. This is not a stable determination of a nonzero odd amplitude. Rather, it shows that the enlarged odd sector is only weakly constrained by the present data.
Numerically, the odd term remains small. In the dip region, the real part of the A3 $C$-odd amplitude does not exceed about $3$--$4\%$ of the real part of the effective $C$-even amplitude.

This smallness is not in contradiction with the fitted value of $\beta_{{\Bbb O}}(0)$. The latter is a residue at $t=0$. At negative $t$ it is multiplied by the trajectory factor
\begin{eqnarray}
\sbar^{\alpha^{\prime}_{{\Bbb O}}t}=\exp(-\alpha^{\prime}_{\Odd}|t|\ln\sbar).
\end{eqnarray}  
For A3, $\alpha'_{\Odd}\simeq0.69~\GeV^{-2}$, and this factor
is only a few $10^{-3}$ at $|t|\simeq0.5~\GeV^2$ in the LHC
energy range. The odd term is therefore strongly reduced in the
dip region. The large errors of the odd residue slopes give the
same message. In A3 the uncertainty in $B_{\Odd}$ is already
large, and in the more flexible A4, B4 and B5 fits it becomes much
larger. Thus the fitted odd residues should not be read as stable
large Odderon amplitudes in the dip region.

The enlarged fits are shown in Table~\ref{tab:a4a5} and in Fig.~\ref{fig:lr_a4a5}. Of course, additional free parameters can lower the value of $\chi^2$. The question is where this freedom is most effective. Adding a fourth Pomeron-like term gives $\tilde\chi^2=1.79$, to be compared with 1.86 in A3. Adding a second Odderon-like term gives instead $\tilde\chi^2=1.84$, which is close to the A3 value. The same comparison in the reweighted ensemble gives 1.44 for B4 and 1.51 for B5, to be compared with 1.50 for B3. Thus the improvement is mainly due to a more flexible effective $C$-even amplitude. It is seen most clearly in the $1.96~\TeV$ and $2.76~\TeV$ contributions. The 7 TeV contribution remains large.

The second Odderon-like term does not give a comparable effect. The new odd parameters are poorly determined. Several of them are compatible with zero, or have large relative uncertainties, including the slopes of the odd residues. Thus the data do not resolve two independent $C$-odd components in the present parametrization.

The same global pattern is found in all model variants. Increasing the freedom in the odd sector does not lead to a substantial change in the curves. The fourth Pomeron-like term gives a better description, but this improvement belongs to the effective $C$-even part of the amplitude. The more visible change in all families comes from replacing the published-error ensemble by the 13 TeV reweighted ensemble. This mainly reduces the 13 TeV contribution, as expected from Eq.~(\ref{eq:berrors}). It does not lead to a comparable improvement at the other energies.

The partial values of $\chi^2_E/N_E$ show the same behavior in all model variants. The 8 TeV data are well described, with $\chi^2_E/N_E$ around $0.7$--$0.8$ in most fits. By contrast, the 7 TeV contribution remains large. The $1.96~\TeV$ and $2.76~\TeV$ samples also give sizeable contributions, although they are improved when the fourth effective Pomeron-like term is included. Since no large physical change is expected between 7 and 8 TeV in this region, this points either to a tension among data sets or to a limitation of the simple effective parametrization.

\section{Conclusions}
\label{sec:conclusion}

We have performed a direct Regge-inspired analysis of elastic $pp$
and $\bar p p$ scattering in the diffractive dip region. The
data with $\sqrt{s}>500~\GeV$ were fitted in the interval $0.2<|t|<1~\GeV^2$.
The amplitude was written as a sum of effective $C$-even and
$C$-odd contributions. Their phases were fixed by the
corresponding signature factors, and were not introduced as
independent fit parameters.

The $C$-even part was first represented by three Pomeron-like terms. This is an effective parametrization in the finite range of $s$ and $t$ considered here. The terms should not be identified with bare Regge poles. They may also represent, in an averaged way, the effect of cuts and absorptive corrections.

The $C$-odd part was tested through a sequence of models. In A1 the Odderon-like trajectory was fixed to $\alpha_{\Odd}(t)=1$. In A2 the slope of the odd residue was allowed to vary independently. In A3 the intercept and the trajectory slope of the odd term were also released. We then made two further checks. In A4 a fourth effective Pomeron-like term was added. In A5 a second Odderon-like term was added. For each case we fitted two data ensembles, the published-error ensemble and the 13 TeV reweighted ensemble.

The result is that a stable nonzero $C$-odd term is not required by the data used here. In A1 the Odderon-like coupling is compatible with zero. In A2 a nonzero odd coupling is obtained, but the fitted residue slope is large, and the odd contribution is strongly suppressed in the dip region. In A3 the additional odd-sector parameters are poorly constrained. The A5/B5 fits give the same message. A second Odderon-like term is not resolved as an independent contribution. The main improvement obtained by enlarging the parametrization comes instead from A4, where a fourth effective Pomeron-like term is introduced. The comparison with A5 shows that the extra freedom is more effective in the $C$-even part of the amplitude. It does not remove the large 7 TeV contribution, and it does not change the conclusion on the odd sector.

The reweighting of the 13 TeV systematic errors is a useful
diagnostic. It lowers the total value of $\chi^2/\nu$, but
mainly by reducing the 13 TeV partial contribution. It does not
remove the larger contributions from the 7 TeV, 2.76 TeV and
1.96 TeV data. The good description of the 8 TeV data, compared
with the poorer description of the 7 TeV data, points to
tensions among data sets.

There is also a more local limitation of the fit. For
$|t|\gtrsim 0.55~\GeV^2$ the description tends to undershoot
both the $pp$ data at $2.76~\TeV$ and the $\bar p p$ data
at $1.96~\TeV$. This defect is not naturally cured by an odd
term alone. Increasing the $C$-odd contribution can raise the
$\bar p p$ cross section, but it lowers the $pp$ one. The
missing contribution in this region should therefore be mainly
$C$-even.

The same point is seen in the comparison of the reduced and more flexible fits. Model A3 gives a smaller total $\chi^2$ than the corresponding reduced fit without a visible odd contribution.
However, the improvement is modest in view of the additional freedom, and the fitted odd amplitude remains small and unstable.
The A5/B5 fits confirm this pattern. The decrease in $\chi^2$ should therefore not be interpreted as a stable determination of the Odderon term.

These conclusions should not be read as evidence that the
Odderon is absent. They refer to the present effective
parametrization and to the data sets used in the fit. The
$C$-even amplitude has considerable freedom and can absorb much
of the dip-region structure. The question tested here is whether, after allowing this freedom, an odd contribution can be resolved in a stable way. In the models considered in this paper the evidence for such a contribution is weak.

This result is complementary to the recent scaling-and-analyticity
analysis of Ref.~\cite{CorralRoyon}. In that approach, the LHC
$pp$ differential cross sections are first represented in terms
of a scaling variable involving $s$ and $t$. The analytic
relation between energy dependence and phase is then imposed
separately for positive and negative signature. The difference
between the general scaling amplitude and the corresponding
positive-signature amplitude gives the negative-signature
part.\footnote{In Ref.~\cite{CorralRoyon} an additional phase $\phi^+$, or equivalently a factor $\exp(i\phi^+)$, is fitted in the second term of the positive-signature amplitude $A^+$. Thus the fitted $A^+$ amplitude is not a pure sum of positive-signature Regge terms in the sense used here, since one of its terms contains an additional phase not generated by the signature factor. This is different from the construction used here, where the phases of all terms are fixed by the corresponding signature factors.} In this way, a small nonzero Odderon amplitude is obtained.

There is no direct contradiction. The scaling analysis tests what follows when the observed scaling behavior and the analytic energy-to-phase relation are imposed. The present analysis tests a different point: whether an odd contribution can be resolved in a stable way in a direct fit with a flexible effective even amplitude. The two procedures therefore probe different aspects of the same problem. In the present fits, the phases are fixed by the signature factors. This makes the separation between the energy dependence and the crossing phase more restrictive.

The interpretation of the odd sector is also limited by the effective parametrization.
With one or two Odderon-like terms, the fit cannot distinguish an Odderon pole from a possible Pomeron-Odderon cut.
A second phenomenological odd term is not the same thing as an explicit cut contribution. A natural extension is
\begin{equation} {\cal F}_{-}(s,t)= {\cal F}_{\Odd}^{\rm pole}(s,t) + {\cal F}_{\Pom\Odd}^{\rm cut}(s,t),
\label{eq:oddcutdecomp}
\end{equation}
where the second term is also $C$-odd. A possible form is
\begin{equation}
{\cal F}_{\Pom\Odd}^{\rm cut}(s,t)= \eta_{\Pom\Odd}(t)\, \beta_{\Pom\Odd}(t)\, {\sbar^{\alpha_{\Pom\Odd}(t)}\over \ln\sbar}.
\label{eq:oddcut}
\end{equation}
This explicit cut term has not been included in the present fits.
It would test whether part of the negative-signature amplitude is missed by the present effective Odderon-like parametrization.

\section*{Acknowledgments}

This research was partially supported by the Conselho Nacional de Desenvolvimento Cient\'ifico e Tecnol\'ogico.


\appendix
\section{The fixed odd trajectory}
\label{app:fixedodd}

The fixed choice
\[
 \alpha_{\Odd}(t)=1
\]
has to be interpreted with some care. If it is taken as a genuine
negative-signature Regge pole with a physical flat trajectory, it would lead to the singular behavior discussed by Petrov and
Tkachenko~\cite{PetrovTkachenko}. This was the basis of their
criticism of the flat-Odderon form\footnote{The term ``Flat Odderon'' was introduced by Petrov and Tkachenko~\cite{PetrovTkachenko} to denote the version of
the model of Ref.~\cite{LunaRyskinKhoze2024} in which the trajectory
of the Odderon-like contribution is fixed to
$\alpha_{\Odd}(t)=1$. We use the term here only in this descriptive
sense.} used in
Ref.~\cite{LunaRyskinKhoze2024}. The criticism, however, rests on
identifying the phenomenological odd term with a complete Regge
pole in the crossed channel. This is not the construction used in
Ref.~\cite{LunaRyskinKhoze2024}, and it is not the construction
used in the present paper.

In the present analysis the odd term is an effective high-energy
parametrization of the $C$-odd elastic amplitude in a finite
interval of $s$ and $t$. It is not continued to the crossed
channel as a particle Regge trajectory. The reduced form used
below retains the crossing phase, but it does not contain the
full Regge-pole denominator or the crossed-channel vertex. This
is the point which removes the singularity.

The argument of Petrov and Tkachenko starts from the standard
negative-signature Regge-pole expression for an Odderon Born
amplitude. In a compact notation it could be written as
\begin{eqnarray}
{\cal O}(s,t) &=& \pi\alpha^{\prime}_{\Odd}(t)
\left[\alpha_{\Odd}(t)+{1\over2}\right]
\left\{ i + \tan\left[{\pi\alpha_{\Odd}(t)\over2}\right] \right\}
\nonumber \\
&\times& \Gamma_{\alpha_{\Odd}(t)}^2(t)
P_{\alpha_{\Odd}(t)}(-z_t),
\label{eq:petrov-pole}
\end{eqnarray}
where $\Gamma_{\alpha_{\Odd}(t)}(t)$ is the crossed-channel
vertex, $P_{\alpha_{\Odd}(t)}$ is the Legendre function, and
\[
 z_t=1-{2s\over 4m_J^2-t}.
\]
The relevant factor is
\begin{equation}
\tan\left[{\pi\alpha_{\Odd}(t)\over2}\right].
\label{eq:petrov-tan}
\end{equation}
If the flat trajectory is approached by writing
\[
 \alpha_{\Odd}(t)=1+\alpha'_{\Odd}t,
\]
then, for small $|\alpha'_{\Odd}t|$,
\begin{equation}
\tan\left[
{\pi\over2}+{\pi\alpha'_{\Odd}t\over2}
\right]
=
-\cot\left({\pi\alpha'_{\Odd}t\over2}\right)
\simeq
-{2\over \pi\alpha'_{\Odd}t}.
\label{eq:tan-limit}
\end{equation}
The factor $\alpha'_{\Odd}$ in Eq.~(\ref{eq:petrov-pole}) is
then cancelled. Up to inessential numerical factors, the real
part of the pole term behaves as
\begin{equation}
{\cal O}(s,t)
\sim
{\Gamma_1^2(t)P_1(-z_t)\over -t}.
\label{eq:petrov-massless-pole}
\end{equation}
Thus, if $\alpha_{\Odd}(t)=1+\alpha'_{\Odd}t$ is interpreted as
a physical flat Regge trajectory, the crossed channel contains a
massless vector pole. To avoid this pole the vertex must vanish
at $t=0$, for example as
\begin{equation}
\Gamma_1^2(t)
\sim
{\rm const.}\,t^{2N},
\qquad N\geq 1.
\label{eq:vertex-zero}
\end{equation}

This chain of reasoning applies to the complete Regge-pole
expression in Eq.~(\ref{eq:petrov-pole}). It does not apply to
the reduced phenomenological amplitude used here. In both
Ref.~\cite{LunaRyskinKhoze2024} and the present paper the odd
term is a high-energy contribution to the elastic amplitude. The
complete Regge-pole expression, with its tangent factor and with
the crossed-channel vertex
$\Gamma_{\alpha_{\Odd}(t)}(t)$, is not used. What is retained is
only the crossing phase.

In the present parametrization the odd contribution is given by
Eq.~(\ref{eq:oddterm}), with the signature factor
(\ref{eq:sigO}). Thus, for model A1,
\begin{equation}
\eta_{\Odd}
=
-i\exp\left(-{i\pi\over2}\right)
=
-1,
\label{eq:a5-phase}
\end{equation}
and therefore
\begin{equation}
{\cal F}_{\Odd}^{\rm A1}(s,t)
=
-\beta_{\Odd}(t)
\left({s\over s_0}\right).
\label{eq:a5-odd-term}
\end{equation}
There is no factor
$\tan[\pi\alpha_{\Odd}(t)/2]$, and no $1/t$ singularity is
generated. The residue $\beta_{\Odd}(t)$ in
Eq.~(\ref{eq:oddterm}) is not the crossed-channel vertex
$\Gamma_{\alpha_{\Odd}(t)}^2(t)$ of
Eq.~(\ref{eq:petrov-pole}). It is an effective residue fitted in
a finite interval of $s$ and $t$.

The same distinction also applies to the phenomenological
construction used in Ref.~\cite{LunaRyskinKhoze2024}. There the
Odderon opacity was obtained from the input amplitude by a
Fourier-Bessel transform,
\begin{equation}
\Omega_{\Odd}(s,b)
=
{2\over s}
\int_0^\infty q\,dq\,J_0(bq)\,
F_{\Odd}^N(s,t),
\qquad t=-q^2 .
\label{eq:opacity-odd}
\end{equation}
The input amplitude was of the form
\begin{equation}
F_{\Odd}^N(s,t)
=
\beta_{\Odd}^2(t)\,
\eta_{\Odd}(t)
\left({s\over s_0}\right)^{\alpha_{\Odd}(t)},
\end{equation}
with the same reduced odd-signature factor as in
Eq.~(\ref{eq:sigO}). For $\alpha_{\Odd}(t)=1$, and for an
exponential residue
\begin{equation}
\beta_{\Odd}(t)
=
\beta_{\Odd}(0)\exp(Dt/2),
\label{eq:old-residue}
\end{equation}
one has, near $q=0$,
\begin{equation}
F_{\Odd}^N(s,-q^2)
\simeq
-\beta_{\Odd}^2(0)
\left({s\over s_0}\right).
\label{eq:finite-input}
\end{equation}
The small-$q$ part of Eq.~(\ref{eq:opacity-odd}) therefore
behaves as
\begin{equation}
q\,dq\,F_{\Odd}^N(s,-q^2)
\sim q\,dq ,
\label{eq:finite-opacity}
\end{equation}
and is regular. By contrast, the literal pole expression
(\ref{eq:petrov-massless-pole}) would give
\begin{equation}
F_{\Odd}^N(s,-q^2)
\sim {1\over q^2},
\label{eq:singular-input}
\end{equation}
so that the same transform would contain
\begin{equation}
q\,dq\,F_{\Odd}^N(s,-q^2)
\sim {dq\over q}.
\label{eq:log-div}
\end{equation}
The logarithmic divergence discussed in
Ref.~\cite{PetrovTkachenko} is therefore a consequence of using
the complete Regge-pole expression (\ref{eq:petrov-pole}) with a
literally flat trajectory. It is not a property of the reduced
phenomenological amplitude used in
Ref.~\cite{LunaRyskinKhoze2024}.

In the present paper the situation is still simpler. We do not
eikonalize the odd term. We fit the elastic amplitude directly in
the finite dip-region interval $0.2<|t|<1~{\rm GeV}^2$.
The parametrization is not continued to $t=0$, and it is not
interpreted as defining a crossed-channel particle trajectory.
The choice $\alpha_{\Odd}(t)=1$ in model A1 is only a minimal
phenomenological test of a nearly energy-independent effective
$C$-odd contribution in this interval. The more flexible model
A3, Eq.~(\ref{eq:a8traj}), is included to check whether the
conclusions depend on this restrictive choice.


\end{document}